# G-SEED: A Spatio-temporal Encoding Framework for Forest and Grassland Data Based on GeoSOT

Ouyang Xuan [1, a], Yu Xinwen[1, b, *], Chen Yan[1, c], Deng Guang[1, d]
and Liu Xuanxin [1, e]

[1]Institute of Forest Resource Information Techniques, Chinese Academy of Forestry.

[a]daylilyau@ifrit.ac.cn, [b]yuxinwen@ifrit.ac.cn, [c]chenyan@ifrit.ac.cn, [d]dengg@ifrit.ac.cn,
[e]liuxx@ifrit.ac.cn

**Abstract.** In recent years, the rapid development of remote sensing, Unmanned Aerial Vehicles, and IoT technologies has led to an explosive growth in spatio-temporal forest and grassland data, which are increasingly multimodal, heterogeneous, and subject to continuous updates. However, existing Geographic Information Systems (GIS)-based systems struggle to integrate and manage of such large-scale and diverse data sources. To address these challenges, this paper proposes G-SEED (GeoSOT-based Scalable Encoding and Extraction for Forest and Grassland Spatio-temporal Data), a unified encoding and management framework based on the hierarchical GeoSOT (Geographical coordinate global Subdivision grid with One dimension integer on $2^n$ tree) grid system. G-SEED integrates spatial, temporal, and type information into a composite code, enabling consistent encoding of both structured and unstructured data, including remote sensing imagery, vector maps, sensor records, documents, and multimedia content. The framework incorporates adaptive grid-level selection, center-cell-based indexing, and full-coverage grid arrays to optimize spatial querying and compression. Through extensive experiments on a real-world dataset from Shennongjia National Park (China), G-SEED demonstrates superior performance in spatial precision control, cross-source consistency, query efficiency, and compression compared to mainstream methods such as Geohash and H3. This study provides a scalable and reusable paradigm for the unified organization of forest and grassland big data, supporting dynamic monitoring and intelligent decision-making in these domains.

**Keywords:** Geospatial data model; GeoSOT; forest and grassland; grid; data query

## 1. Introduction

In recent years, the widespread adoption of remote sensing, the Internet of Things (IoT), and unmanned aerial vehicles (UAVs) technologies has ushered forest and grassland monitoring into the era of "data explosion." In China alone, forest resources generate more than 20 petabytes(PB) of multi-source imagery and ground observation data annually. The data types have expanded beyond optical imagery to include LiDAR point clouds, phenological video streams, and sensor-based time series, exhibiting multi-source heterogeneity, massive growth, and real-time updates [1]. However, due to significant differences in data complexity, origin, timeliness, and descriptive format, the overall efficiency of data utilization remains low, and deep analysis and mining are still lacking. Existing forest and grassland data management systems—primarily based on traditional Geographic Information Systems (GIS)—are increasingly revealing bottlenecks in heterogeneous data integration, incremental updating, and high-concurrency spatial computing. Typical issues include long retrieval times for cross-scale queries, difficulty in version synchronization, and heavy reliance on manual preprocessing for fine-grained analysis. To address these challenges, it is imperative to adopt more efficient technologies that enable the organization and management of diverse forest and grassland data within a unified framework.

In response to these limitations, the research community has begun exploring "unified encoding-driven data organization and management methods," among which the GeoSOT (Geographical coordinate global Subdivision grid with One dimension integer on $2^n$ tree) spatial subdivision system stands out as a representative approach. GeoSOT recursively subdivides the Earth's surface using a $2^n$-based quadtree or octree structure and assigns a unique one-dimensional integer code to each grid unit. This enables three core capabilities: globally unified subdivision, multi-level scalability, and algorithmically computable mapping [2]. GeoSOT has already been widely applied in fields such as remote sensing image management [3], satellite-ground data interaction [4], large-scale remote sensing data indexing and querying [5], and the construction of spatio-temporal cultural tourism data visualization systems [6], demonstrating significant advantages in the unified management of multi-scale and multi-dimensional data.

Against this backdrop, this paper proposes G-SEED (GeoSOT-based Scalable Encoding and Extraction for Forest and Grassland Spatio-temporal Data), a GeoSOT-based spatio-temporal encoding and management framework tailored for forest and grassland data. The framework aims to address the critical challenge of "how to uniformly organize forest and grassland data with varying structures, granularities, and temporal distributions." The main contributions of this study include: (1) constructing a scalable GeoSOT encoding system that defines mapping rules between data types and spatial levels; (2) designing a unified ingestion and integration mechanism for heterogeneous data such as remote sensing imagery, vector layers, monitoring records, textual documents, and multimedia files; (3) validating the framework's effectiveness on real-world datasets in terms of multi-source encoding consistency, adaptive level selection, and code uniqueness; and (4) conducting comparative experiments to evaluate the proposed method's performance in encoding continuity, polygon compression, and retrieval efficiency, in contrast to mainstream alternatives such as Geohash and H3. Through this research, the proposed G-SEED framework aspires to provide a generalized, reusable paradigm for data integration, dynamic analysis, and intelligent decision support in forestry applications, and to lay the foundation for building a spatio-temporal data infrastructure aimed at ecological security and resource protection.

## 2. Related work

In recent years, a variety of innovative approaches have been proposed to address the challenges of integrating, indexing, and querying large-scale, multi-source, and heterogeneous spatio-temporal data in forest and ecological monitoring based on GeoSOT. Among these, solutions utilizing GeoSOT spatio-temporal indexing have attracted considerable attention due to their applicability in the unified management of multimodal data such as LiDAR point clouds, satellite imagery, sensor records, and UAV videos. Qian et al. extended the GeoSOT quadtree spatial grid into the temporal dimension, constructing the GeoSOT ST-index and deploying it within MongoDB [7]. Experimental results showed that this index structure achieved approximately a 40% improvement in range query performance compared to traditional composite spatio-temporal indexes.

Building upon this, Liu et al. proposed a novel spatio-temporal encoding scheme named HGST, which integrates Hilbert curves with the GeoSOT grid [8]. This method leverages the spatial continuity of Hilbert encoding and the hierarchical subdivision of GeoSOT, and implements efficient filtering and refinement processing on the HBase and Spark platforms. Experimental results demonstrated that HGST improved query response speed by 14–35% compared to the GeoSOT ST-index and GeoMesa, showcasing strong scalability. This approach is of direct relevance for real-time event monitoring and trajectory tracking in the forestry and grassland domain.

More recently, Shi et al. introduced the STUCCM model, a unified spatio-temporal coding mechanism that combines longitude, latitude, elevation, and time into a single binary code via bitwise operations, enabling efficient spatio-temporal querying in digital twin city scenarios [9]. This bit-level coding strategy demonstrates exceptional performance in both scalability and precision control, offering a promising architectural reference for future forest and grassland digital twin systems.

In the context of forest resource management, Liu Yongjie developed a unique, multi-scale, discrete spatial coding model based on the globally subdivided GeoSOT grid. This model supports the unified organization, integrated management, and statistical analysis of national forest land data, and an experimental system was constructed to validate its efficiency in data management, transmission, and retrieval access [10]. However, existing applications in the forestry domain remain largely focused on "static" forest land encoding or single-scale statistical analysis. They have yet to fully incorporate the temporal dimension or establish unified strategies for organizing unstructured data such as video and IoT monitoring streams.

In summary, the GeoSOT system continues to demonstrate multi-dimensional advantages in organizing and applying multi-source forest and grassland data. Its hierarchical spatio-temporal indexing structure enables efficient management of heterogeneous resources such as remote sensing imagery, UAV video, and sensor data. Its encoding strategies are continuously evolving, with HGST enhancing spatial continuity and query performance, and STUCCM enabling bit-level precision and parallel scalability. In terms of system deployment, the GeoSOT coding framework is highly compatible with big data processing platforms such as MongoDB, HBase, and Spark, and is well-suited for architectures targeting digital twin environments. The integration and advancement of these technologies offer a solid foundation for overcoming current challenges related to data fusion, delayed monitoring, and insufficient intelligent analysis, thereby supporting the future construction of intelligent monitoring systems and ecological data infrastructures in the forestry domain.

## 3. Method

### 3.1 GeoSOT-Based Spatio-temporal Partitioning Scheme for Forest and Grassland Data

3.1.1 Organization of Forest and Grassland Spatio-temporal Data

Forest and grassland spatio-temporal data can be categorized into two main types: structured and unstructured data. Structured data includes raster datasets from remote sensing (e.g., high-resolution imagery, thumbnails, satellite data), vector data, and ecological monitoring records. These datasets typically follow fixed formats, can be indexed, and are convenient for query operations. In contrast, unstructured data encompasses field survey reports, policy documents, and multimedia content such as videos, audio recordings, and photographs. Due to the lack of a unified organizational structure, unstructured data often requires additional parsing and processing. Structured data is typically stored in relational or spatial databases, supporting SQL and spatial indexing queries; unstructured data is generally housed in document databases or object storage systems and requires information extraction through techniques such as natural language processing (NLP), image recognition, or temporal analysis.

3.1.2 Spatio-temporal Partitioning Scheme for Forest and Grassland Data

To address the challenges of integrating multi-source heterogeneous data in forest and grassland information management, the GeoSOT grid system offers notable advantages over traditional GIS-based approaches. First, GeoSOT assigns a globally unique identifier to each spatial unit via Z-order quaternary encoding, ensuring spatial uniqueness. Second, its multi-level grid subdivision mechanism provides strong adaptability to scale, allowing dynamic matching of spatial resolution based on data precision. Third, GeoSOT offers a unified spatial encoding framework capable of efficiently linking raster data, vector data, and non-spatial attributes, thus laying a foundation for the integrated management and unified retrieval of heterogeneous data.

Forest and grassland spatio-temporal data are uniformly partitioned via the GeoSOT grid system. Each data type is mapped to an appropriate grid level based on its spatial resolution and positioning accuracy, thereby enabling consistent encoding and cross-scale integration. The coarsest spatial resolution (level 7, ~445.3 km grid size) is designed for large-scale ecological analyses and regional climate studies, while the finest resolution (level 24, ~3.9 m grid size) enables detailed environmental monitoring and individual object tracking. This partitioning strategy ensures cross-scale compatibility within a unified encoding system, facilitates efficient data aggregation, and supports spatio-temporal representation from macro-level trends to micro-level dynamics. The specific design of the partitioning scheme is presented in Table 1.

Table 1. GeoSOT-Based Spatio-temporal Partitioning Scheme for Forest and Grassland Data

| Grid Level | Spatial Resolution | Data Type | Application Scenario |
|---|---|---|---|
| Level 7–14 | 3.7–445.3 km | County-level statistical reports, provincial boundary data | Regional resource inventory and evaluation |
| Level 15–17 | 494.7 m – 1.8 km | Medium- to low-resolution remote sensing imagery | Forest cover change monitoring |
| Level 18–21 | 30.9 – 247.4 m | High-resolution imagery, forest compartment boundaries | Plot-level area estimation |
| Level 22–24 | 3.9 – 15.5 m | UAV-based LiDAR data, ground-based sensor data | Individual tree identification |

In terms of remote sensing imagery, high-resolution data are mapped to GeoSOT grid levels 22–24 to preserve detailed structural features, while medium- and low-resolution imagery are mapped to levels 15–17, making them suitable for regional-scale analysis and change detection.

For vector boundary data, forest compartment polygons are allocated to grid levels 18–21. Linear features such as roads are encoded in segments based on line length, and grid intersection nodes are aligned with grid centers to ensure consistency in encoding.

Sensor monitoring data employ a dynamic granularity matching strategy: meteorological stations are mapped to level 17 grids with hourly resolution; phenological cameras are assigned to level 21 grids with daily resolution; soil moisture monitoring uses level 22 grids with acquisition-frequency granularity; and wildlife tracking is encoded at level 24 with a 10-minute interval, supporting high-frequency, fine-scale dynamic monitoring tasks.

Textual data are mapped to grid levels 7–14 based on administrative hierarchy. Multimedia data extract EXIF metadata or spatial location information and are mapped to levels 21–24; if GPS metadata are unavailable, such data are instead mapped to levels 7–14 based on the affiliated institution's location.

Through this mechanism, forest and grassland spatio-temporal data achieve spatial consistency and hierarchical compatibility across multi-source and multimodal datasets within a unified GeoSOT framework. This effectively supports cross-scale integration and high-performance spatial retrieval. The mapping and organization of forest and grassland data within the GeoSOT grid system are illustrated in Figure 1.

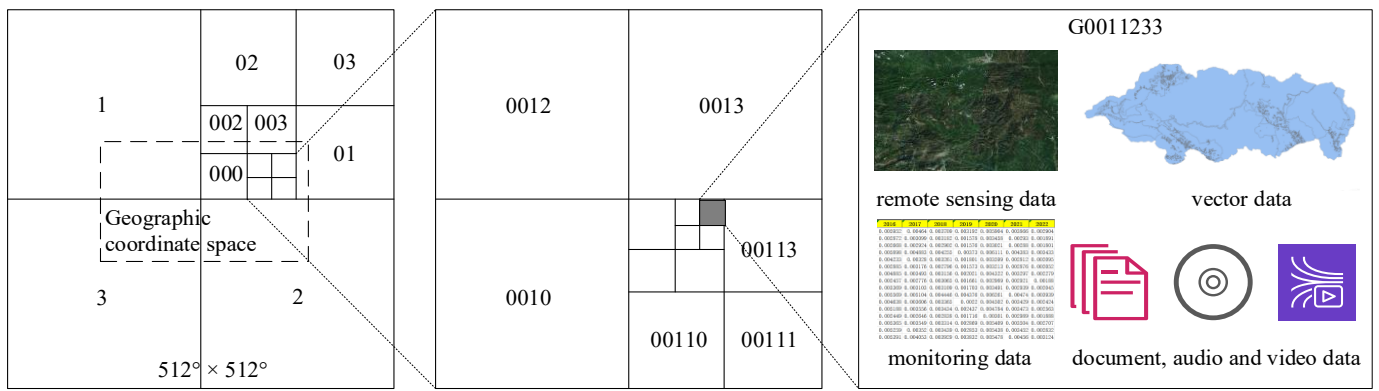

Fig 1. GeoSOT-Based Organization of Forest and Grassland Spatio-temporal Data

## 3.2 GeoSOT-Based Encoding for Forest and Grassland Spatio-temporal Data

### 3.2.1 Encoding Structure

To enable unified identification and efficient management of forest and grassland spatio-temporal data, this study designs a composite encoding structure that integrates spatial, temporal, and type information. The complete encoding format is as follows:

GeoSOT Code + Timestamp + Type Identifier

GeoSOT Code: A one-dimensional quaternary code in which each digit corresponds to a hierarchical grid cell, arranged according to a Z-order curve. The code begins with the prefix "G"; for example, "G001123332013230" denotes a level-15 spatial grid cell, enabling precise representation and localization of spatial positions.

Timestamp: To support time series analysis and data versioning, a date field is included as a temporal identifier in the encoding structure. It adopts the standard "YYYYMMDD" format and can be extended to include hour (HH) or minute (mm) granularity if needed. This design is especially suitable for high-frequency monitoring data and time-series remote sensing imagery that require strong temporal resolution.

Type Identifier: In response to the diverse data types present in forest and grassland spatio-temporal datasets, a concise and unified type code system is established. The specific design is shown in Table 2.

Table 2. Type Identifier Design

| Type Code | File Formats | Data Type |
|-----------|--------------|-----------|
| RAS | .tif / .tiff, .nc | Remote Sensing |
| VEC | .shp, .geojson, .gpkg, .kml | Vector Data |
| TB | .csv, .xls | Tabular Data |
| DOC | .doc, .pdf, .txt | Documents |
| IMG | .jpg, .png | Images |
| AUD | .wav, .flac, .mp3, .aac, .ogg | Audio |
| VEO | .mp3, .mp4, .avi | Video |

### 3.2.2 Optimization Strategy

To address the issues of spatial indexing redundancy and decreased retrieval efficiency caused by large imagery patches, elongated roads, or river networks spanning multiple GeoSOT grids, this study proposes an optimization strategy integrated into the G-SEED (GeoSOT-based Scalable Encoding and Extraction for Forest and Grassland Spatio-temporal Data) framework. The strategy incorporates three core mechanisms: adaptive grid-level selection, central grid localization, and coverage grid recording. The details are as follows:

Adaptive Grid-Level Selection: Depending on the user's spatial query scope, the most suitable GeoSOT level is automatically selected for filtering. Lower-level grids are used for rapid positioning in large-scale queries, while higher-precision grids are selected for small-scale queries to ensure accuracy—achieving a dynamic balance between spatial resolution and system overhead.

Central Grid Localization: For objects spanning multiple grids, the grid cell containing the geometric center is selected as the primary retrieval keyword. This leverages the prefix matchability of GeoSOT codes to enhance retrieval speed.

Coverage Grid Recording: All GeoSOT grid codes crossed by the target object are fully recorded within the data entry to support more precise spatial intersection queries and position reconstruction.

This strategy not only extends existing approaches that rely solely on single-grid or fixed-level encodings but also enhances the accurate representation and retrieval of complex, cross-grid spatial objects. It exhibits strong compatibility and scalability, making it well-suited for managing large-scale spatial raster and vector datasets. The optimized GeoSOT-based spatio-temporal data encoding framework for forest and grassland applications (G-SEED) is illustrated in Figure 2, with the corresponding pseudocode logic provided below the figure.

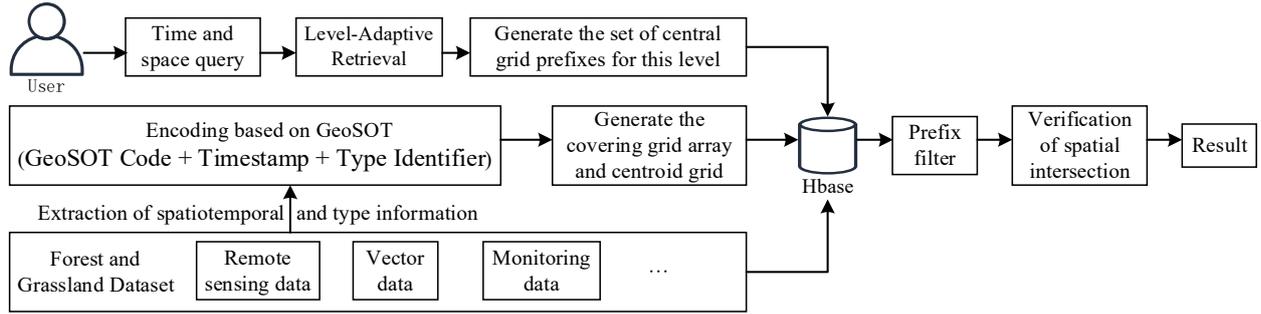

Fig.2 G-SEED: GeoSOT-based Scalable Encoding and Extraction for Forest and Grassland Spatio-temporal Data

```
Input:
    F – file path
    G – geometry (point, line, polygon, raster extent, etc.)
    T – timestamp (YYYYMMDD[HHmm])
    M – metadata (resolution, source type, GPS tag, etc.)
Output:
    PK – composite primary key: GeoID | Timestamp | TypeCode
    GeoID – grid code of central cell (Z-order encoded)
    Covers – list of covering grids (optional)
    L – selected grid level
    meta – retained metadata
# --- Step 1: Type Identification ---
1    ext  ←  GetFileExtension(F)
2    typ  ←  DetectType(ext)   # 'RAS', 'VEC', 'TB', 'DOC', 'IMG', 'AUD', 'VEO'
# --- Step 2: Grid Level Selection (Adaptive) ---
3    if typ == 'RAS' then
4        L  ←  LevelByResolution(M.res)              # Levels 15–24
5    elif typ == 'VEC' then
6        G  ←  DouglasPeucker(G, ε)                  # Simplify geometry
7        L  ←  LevelBySize(G)                        # Levels 18–21
8    elif typ in ['IMG', 'VEO', 'AUD'] then
9        L  ←  (M.gps ? 22 : 10)                     # GPS-tagged or default
10   elif typ in ['TB', 'DOC'] then
11       L  ←  LevelByAdminRegion(M.admin_level)     # Levels 7–14
12   else
13       L  ←  12                                    # Fallback level
# --- Step 3: Central Grid Encoding ---
14   (lon, lat)  ←  Centroid(G)
15   GeoID  ←  ZEncode(lon, lat, L)                  # Generate GeoSOT grid code
# --- Step 4: Covering Grids (Optional for large or linear features) ---
16   Covers  ←  []
17   if Size(G) > GridSize(L) then
18       Covers  ←  CoverGeometry(G, L)              # Record full coverage
# --- Step 5: Composite Key Construction ---
19   TS  ←  FormatTimestamp(T, "YYYYMMDD[HHmm]")
20   PK  ←  GeoID | "|" | TS | "|" | typ
# --- Step 6: Return Encoded Result ---
21   return { PK, level: L, center: GeoID, covers: Covers, meta: M }
```

# 4. Experiments and Results

## 4.1 Dataset

This study employs a dataset collected and curated over years by our research team, sourced from the Shennongjia National Park. The dataset includes five major categories of data: remote sensing imagery, vector boundaries, ecological monitoring records, survey documents, and multimedia observations. In total, it comprises approximately 14,570 files, with an overall data volume of approximately 101.2 GB. The core of the dataset consists of high-resolution remote sensing imagery, complemented by vector data representing forest plots and administrative boundaries. It also integrates long-term ecological monitoring tables and survey reports (structured and semi-structured documents), along with multimedia data such as images and videos collected from field-deployed UAVs and infrared cameras. Additionally, the dataset includes audio recordings for ecoacoustic and environmental monitoring, further enriching its spatio-temporal informational dimensions.

Located in western Hubei Province, Shennongjia National Park is one of China's 14 biodiversity hotspots of global significance, with a forest coverage exceeding 96%. The region features significant altitudinal gradients and diverse climatic types, making it a long-standing "natural laboratory" for protected area monitoring and ecological process research [11]. Leveraging this multimodal dataset provides strong support for ecological health assessment, vegetation classification, and process tracking, and offers a robust data foundation for forest ecosystem research and regional ecological management.

## 4.2 G-SEED Encoding Performance Experiments

### 4.2.1 Multi-Level Spatio-temporal Encoding Performance of G-SEED

To assess the performance of G-SEED in spatial granularity control and spatio-temporal encoding efficiency, a multi-level encoding experiment was conducted. All spatial points from the heterogeneous forest and grassland dataset were extracted and encoded at four different levels (15, 17, 19, and 21). Key indicators—such as average code length, redundancy rate, number of unique codes, storage size, and encoding time—were recorded to compare the encoding outcomes across different levels. Results are presented in Table 3.

Table 3. Performance results of G-SEED multi-level encoding

| Level | Avg Length | Repeat Rate | Unique Codes | Storage (bytes) | Time (s) |
|-------|-----------|-------------|--------------|-----------------|----------|
| 15 | 22 | 0.8792 | 2457 | 345746.0 | 0.1122 |
| 17 | 25 | 0.6824 | 6460 | 406760.0 | 0.1141 |
| 19 | 27 | 0.3550 | 13119 | 447436.0 | 0.1200 |
| 21 | 29 | 0.1237 | 17823 | 488112.0 | 0.1249 |

The results show that higher encoding levels improve spatial resolution: redundancy drops from 87.9% (level 15) to 12.4% (level 21), while the number of unique codes rises significantly. Higher encoding levels reduce redundancy by assigning more unique codes to finer spatial units. Meanwhile, encoding time remains stable (~0.12 s on average). This demonstrates that the proposed scheme supports flexible spatial precision control without incurring substantial computational overhead, achieving a favorable balance between encoding compression and indexing efficiency. It is thus well-suited for multi-scale and multi-task spatial data processing scenarios in the forestry domain.

### 4.2.2 Adaptive Level Selection Experiment

To evaluate G-SEED's adaptability to datasets with varying spatial scales, an adaptive level selection experiment was designed. By calculating the spatial bounding area of all data points, the approximate area was estimated, and an appropriate encoding level was automatically selected based on a predefined area-to-level mapping rule. Encoding was then performed, and metrics such as redundancy rate, number of unique codes, average code length, and time consumption were recorded to assess adaptability. The results are shown in Table 4.

Table 4. Results of Adaptive Level Selection Experiment

| Level | Area (km²) | Avg Length | Repeat Rate | Unique Codes | Time(s) |
|-------|-----------|-----------|-------------|--------------|---------|
| 15 | 2.1 | 22 | 0.8792 | 2457.0 | 0.1119 |

The experiment showed that within a spatial extent of approximately 2.1 km², G-SEED automatically selected level 15 for encoding. The results yielded a redundancy rate of 87.9%, an average code length of 22, and an encoding time of just 0.11 seconds, demonstrating both strong spatial compression and execution efficiency. This experiment further verifies that G-SEED can intelligently adapt to spatial distribution characteristics, selecting the optimal granularity level to achieve a practical balance between representational fidelity and storage efficiency—suitable for dynamic, multi-scale forest and grassland data processing.

4.2.3 Consistency Verification for Heterogeneous Data Encoding

To evaluate the adaptability and consistency of G-SEED in handling heterogeneous forest and grassland data, a multi-source data fusion consistency experiment was conducted. The experiment traversed the entire dataset, including various spatial data types such as points, lines, polygons, and raster images. For each data type, spatial coordinates were extracted and encoded uniformly at level 15. For each file, total number of codes, number of unique codes, and redundancy rate were calculated to assess performance across data types. Results are summarized in Table 5.

Table 5. Results of Consistency Verification for Heterogeneous Data Encoding

| Data Type | Average Redundancy Rate |
|-----------|-------------------------|
| Point | 0.0000 |
| Line | 0.4327 |
| Polygon | 0.9516 |
| Raster | 0.0000 |

The results show that point data and remote sensing rasters achieved a 0% redundancy rate, indicating excellent code uniqueness. Linear data (e.g., roads and boundaries) exhibited moderate redundancy (~43.3%), reflecting their inherent tendency to span multiple grids. Polygonal features (e.g., land use parcels) showed a high redundancy rate of 95.2%, demonstrating G-SEED's potential for compressing and representing large areal features. Overall, G-SEED, through its unified encoding structure, effectively covers a wide range of spatial data types and formats, enabling consistent and integrated management of heterogeneous sources with strong generalizability and expressive consistency.

## 4.3 Comparison with Traditional Methods

Geohash is a latitude-longitude encoding scheme based on the Z-order curve. It recursively partitions the Earth's surface into rectangular grids and encodes each grid location using a Base32 string. Its support for variable precision and prefix matching has made it widely used in spatial indexing and location clustering tasks. However, it suffers from issues such as boundary fragmentation and spatial discontinuity between adjacent cells [12].

H3 (Hexagonal Hierarchical Spatial Index), developed by Uber, adopts a spherical hexagonal tessellation combined with a multi-level indexing scheme. It employs integer-based encoding to achieve efficient spatial representation and neighborhood searching, and has been widely adopted in large-scale geospatial data processing and visualization [13].

Since GeoSOT also employs a hierarchical spatial division for encoding—featuring multi-scale representation, prefix consistency, and spatial compression capability—this study selects Geohash and H3 as benchmark methods to comprehensively evaluate the performance advantages and applicability of G-SEED in forest and grassland spatio-temporal data organization.

### 4.3.1 Spatial Continuity Test

To assess the ability of different spatial encoding methods to preserve spatial adjacency, a spatial continuity experiment was designed. A group of adjacent points was selected from the dataset and encoded using G-SEED, Geohash, and H3, respectively. The prefix portion of each code was extracted, and the proportion of points with matching prefixes was calculated to determine prefix consistency, which reflects how well spatial proximity is preserved in the encoding structure. Results are shown in Table 6.

Table 6. Results of Spatial Continuity Test

| Encoding Method | Prefix Consistency |
|---|---|
| G-SEED | 0.82 |
| Geohash | 0.02 |
| H3 | 0.42 |

G-SEED achieves a prefix consistency of 0.82, significantly outperforming H3 (0.42) and Geohash (0.02). This highlights G-SEED's superior ability to preserve spatial adjacency. High prefix consistency makes it more suitable for tasks such as spatial clustering, index construction, and prefix-based query, showing enhanced spatial locality—an advantage for managing large-scale forest and grassland spatial datasets.

### 4.3.2 Polygon Coverage Compression Capability

To compare the spatial compression capabilities of different encoding methods in handling large polygonal areas, a multi-grid coverage experiment was designed. Using the extracted point data, a minimum bounding rectangle was constructed to simulate a typical land parcel region. Within this region, spatial points were evenly sampled and encoded using G-SEED, Geohash, and H3, respectively. The number of unique grid cells required to cover the region was counted to assess spatial partitioning and compression efficiency. The results are presented in Table 7.

Table 7.Polygon Coverage Compression Comparison

| Encoding Method | Number of Grid Cells |
|---|---|
| G-SEED | 121009 |
| Geohash | 553750 |
| H3 | 75960 |

The coverage tests show that G-SEED achieves a good balance between spatial regularity and compression efficiency. It uses only 121,009 grid cells to represent the test polygons, significantly fewer than Geohash (553,750). Although H3 is more compact (75,960 cells), G-SEED retains structural advantages such as better alignment with raster data and compatibility with forestry workflows. These features make it well-suited for encoding and managing large-scale forest areas and remote sensing imagery.

### 4.3.3 Encoding and Query Efficiency Evaluation

To thoroughly evaluate the encoding efficiency and query performance of G-SEED, Geohash, and H3, an encoding-query efficiency experiment was conducted. Approximately 20,000 valid spatial point records were extracted from the dataset and encoded using each of the three methods. Key metrics such as encoding time, average code length, total storage, and redundancy rate were recorded. A simulated query task was then executed by defining a unified spatial prefix range for each encoding method, measuring the number of records matched and query execution time. The results are provided in Table 8.

Table 8. Encoding and Query Efficiency Evaluation Results

| Method | Encoding time(ms) | Avg length | Query prefix | Query matched | Query time(ms) |
|---|---|---|---|---|---|
| G-SEED | 3.289 | 17 | G0011233 | 18827 | 0.177 |
| Geohash | 4.911 | 8 | Wmw2x | 47 | 0.233 |
| H3 | 2.747 | 15 | H3_31 | 12978 | 0.321 |

The results demonstrate that G-SEED outperforms the other methods in terms of both encoding and query efficiency. Despite having a longer average code length (17 characters), it achieves precise matching for 18,827 records, significantly surpassing Geohash and H3. Notably, it also achieves the lowest query latency at 0.177 ms, highlighting its suitability for high-precision spatial clipping and low-latency retrieval scenarios—ideal for large-scale forest and grassland spatio-temporal data querying and positioning.

## 5. Summary

This paper proposes a GeoSOT-based encoding and management framework (G-SEED) for heterogeneous spatio-temporal data in forest and grassland environments, addressing current structural and performance limitations in data organization, compression, and efficient retrieval. Based on a unified GeoSOT grid system, the framework constructs a compound encoding structure that integrates spatial, temporal, and type information, supporting consistent representation and indexing for heterogeneous data types such as points, lines, polygons, rasters, documents, and multimedia.

Three optimization mechanisms were designed: hierarchical adaptive retrieval, central grid indexing, and multi-grid coverage recording, which together enhance spatial continuity and polygon compression efficiency.

Experimental results on a real-world dataset show that G-SEED achieves superior flexibility, consistency, and computational performance. Compared to mainstream methods such as Geohash and H3, it offers significant advantages in spatial accuracy, prefix matching precision, and query latency. This study provides a scalable and deployable encoding paradigm for big data management, real-time monitoring, and intelligent decision-making in forestry, laying a solid foundation for future development of ecological digital infrastructures and digital twin systems.

## Acknowledgment

This research was funded by the Chinese Academy of Forestry through the Fundamental Research Funds for Central Non-profit Research Institutions (Grant No. CAFYBB2022SY033), as part of the project "Spatio-temporal Data Organization Model and Application for Forest and Grassland Based on GeoSOT Encoding."